\documentclass{article} 
\usepackage{iclr2025_conference,times}

\usepackage{hyperref}
\usepackage{url}

\usepackage{latexsym}
\usepackage{multirow}
\usepackage{subfigure}
\usepackage{amsmath}
\usepackage{booktabs}
\usepackage{amssymb}
\usepackage{microtype}
\usepackage{balance}
\usepackage{inconsolata}
\usepackage{adjustbox}
\usepackage{graphicx}

\usepackage[utf8]{inputenc} 
\usepackage[T1]{fontenc}    
\usepackage{hyperref}       
\usepackage{url}            
\usepackage{booktabs}       
\usepackage{amsfonts}       
\usepackage{nicefrac}       
\usepackage{microtype}      
\usepackage{xcolor}         

\usepackage{multirow}
\usepackage{subfigure}

\usepackage[english]{babel}
\usepackage{moresize}
\usepackage{amsmath}
\usepackage{algorithmic}
\usepackage{balance}
\usepackage{comment}
\usepackage{paralist}
\usepackage{bm}
\usepackage{pgfplots}
\usetikzlibrary{pgfplots.dateplot}

\usepackage{flushend}
\usepackage[english]{babel}
\usepackage{graphicx}

\usepackage{amssymb}
\usepackage{amsfonts}
\usepackage{url}
\usepackage{bbm}
\usepackage{longtable}
\usepackage{rotating}
\usepackage{multirow}
\usepackage{mathrsfs}
\usepackage{enumitem}
\usepackage[linesnumbered,algoruled,boxed,lined]{algorithm2e}
\usepackage{adjustbox}
\usepackage{hyperref}
\usepackage{pgfplots}
\usetikzlibrary{pgfplots.dateplot}
\usepackage{filecontents}
\usepackage{wrapfig}
\usepackage{caption}

\newcommand{\eg}{\textit{e}.\textit{g}.\xspace}

\def\model{RecLM\xspace}

\title{RecLM: Recommendation Instruction Tuning}

\author{Yangqin Jiang$^{1}$, Yuhao Yang$^{1}$, Lianghao Xia$^{1}$, Da Luo$^{2}$, Kangyi Lin$^{2}$, \textbf{Chao Huang}$^{1}$\noindent\thanks{Chao Huang is the corresponding author.} \\
University of Hong Kong$^1$, Tencent$^2$ \\
\texttt{\small \{mrjiangyq99,chaohuang75\}@gmail.com\; yuhao-yang@outlook.com\; aka\_xia@foxmail.com}
}

\begin{document}

\maketitle

\begin{abstract}
Modern recommender systems aim to deeply understand users' complex preferences through their past interactions. While deep collaborative filtering approaches using Graph Neural Networks (GNNs) excel at capturing user-item relationships, their effectiveness is limited when handling sparse data or zero-shot scenarios, primarily due to constraints in ID-based embedding functions. To address these challenges, we propose a model-agnostic recommendation instruction-tuning paradigm that seamlessly integrates large language models with collaborative filtering. Our proposed \underline{Rec}ommendation \underline{L}anguage \underline{M}odel (\model) enhances the capture of user preference diversity through a carefully designed reinforcement learning reward function that facilitates self-augmentation of language models. Comprehensive evaluations demonstrate significant advantages of our approach across various settings, and its plug-and-play compatibility with state-of-the-art recommender systems results in notable performance enhancements. The implementation of our \model\ framework is publicly available at: \textcolor{blue}{\url{https://github.com/HKUDS/RecLM}}.
\end{abstract}

\section{Introduction}
\label{sec:intro}
Recommendation systems serve as essential components of modern web applications, helping users navigate through the vast digital information landscape. These systems provide personalized suggestions across diverse platforms, including product recommendations on e-commerce platforms~\citep{wang2020time, wu2018turning}, content discovery on social networking sites~\citep{jamali2010matrix, zhang2021understanding}, and viewer-tailored suggestions on video sharing platforms~\citep{zhan2022deconfounding,jiang2024diffmm}. At the core of these systems lies Collaborative Filtering (CF), a widely adopted approach that harnesses collective user preferences to generate personalized recommendations.

Current recommender systems predominantly operate within the user/item ID paradigm, where training data consists primarily of mapped user and item indices. While this approach has driven significant advancements in recommendation technology, particularly in data-rich scenarios~\citep{yao2021self,yuan2023go}, it faces several fundamental limitations. Most notably, these systems encounter substantial challenges in cold-start scenarios and struggle to generalize effectively in zero-shot learning situations. The inherent dependency on ID-based representations becomes particularly problematic in completely cold-start settings, where systems fail to generate meaningful representations for new items, ultimately compromising their ability to deliver accurate recommendations.

Addressing the fundamental cold-start challenge in ID-based recommendation systems requires moving beyond traditional ID-based embeddings to leverage external features (\eg, textual or visual information) for generating user and item representations. However, this promising approach encounters significant practical challenges in real-world applications, primarily stemming from two critical issues: data incompleteness and quality concerns. On the incompleteness front, users frequently withhold personal information due to privacy concerns, leading to substantial gaps in external features. Simultaneously, existing data often suffers from quality issues, manifesting as misleading tags, irrelevant product specifications, or unreliable item descriptions. These challenges collectively underscore the critical need for robust methods to extract accurate, relevant, and high-quality features from inherently noisy and incomplete side information—a prerequisite for achieving effective recommendation generalization in data-scarce scenarios.

\textbf{Contribution.} The exceptional generalization and reasoning capabilities of Large Language Models (LLMs) motivate a novel solution to cold-start recommendation challenges through specialized profiling systems based on external side information. This approach centers on developing effective language models customized for recommendation tasks, addressing two critical challenges: (i) designing robust mechanisms for LLMs to generate accurate profile representations that capture essential recommendation characteristics, especially for users or items lacking historical interactions, and (ii) developing techniques for LLMs to distill high-quality profiles from noisy feature sets while preserving the integrity of user-item interaction patterns and behavioral context.


To address these challenges, we propose \model, a novel recommendation instruction tuning paradigm that revolutionizes how LLMs understand behavioral contexts in recommender systems. While LLMs demonstrate superior natural language processing capabilities, they fundamentally lack the ability to model complex user-item interactions and behavioral preferences. Our approach tackles this limitation through two key technical innovations: (1) a seamless integration mechanism that fuses external user-item features with collaborative interaction patterns through specialized instruction tuning, enabling effective cold-start profiling without direct supervision signals, and (2) a reinforcement learning-based personalized feature enhancement framework that systematically reduces noise and mitigates over-smoothing effects inherent in collaborative filtering. This model-agnostic framework can be plugged into existing recommender systems to significantly enhance their generalization capacity, particularly in data-scarce scenarios. Our main contributions are as follows:

\begin{itemize}[leftmargin=*]
    \item \textbf{Model-Agnostic Framework}. We introduce a model-agnostic instruction tuning framework \model. It can be seamlessly integrated into existing recommender systems as a plug-and-play component, significantly enhancing their generalization capacity in scenarios with data scarcity.


    \item \textbf{Enhancing Profiling System}. We seamlessly integrate large language models with collaborative filtering to enhance user profiling and representation, particularly in cold-start scenarios, where current methods often struggle. Additionally, our approach employs reinforcement learning to refine profile quality, effectively addressing challenges associated with data noise and over-smoothing.
    
    \item \textbf{Comprehensive Evaluation}. We integrate \model\ with a range of state-of-the-art recommenders to assess the effectiveness of our approach across various settings. This includes conducting ablation studies and efficiency evaluations. Additionally, we carry out extensive experiments in real-world industrial recommendation scenarios, demonstrating the practicality and scalability of \model.
    
\end{itemize}
\section{Methodology}
\label{sec:solution}

\subsection{ID-based Collaborative Filtering}

In the ID-based collaborative filtering (CF) paradigm, the primary goal is to optimize the ID embeddings of users and items. This optimization aims to accurately capture and represent user preferences for items, while considering the interaction patterns of users and items that are similar. Formally, we have a set of users denoted as $\mathcal{U} = \{u_1, \cdots, u_{I}\}$, and a set of items denoted as $\mathcal{V} = \{v_1, \cdots, v_{J}\}$. Each user and item is assigned initial ID embeddings, represented as $\mathbf{x}_{u}$ and $\mathbf{x}_{v}\in\mathbb{R}^d$ respectively. The objective is to obtain optimized user and item representations, denoted as $\mathbf{e}_u, \mathbf{e}_v\in\mathbb{R}^d$, through a recommender model $\mathcal{R}(\mathbf{x}_{u}, \mathbf{x}_{v})$. This model aims to maximize the posterior distribution $p(\mathbf{e}|\mathcal{X}) \propto p(\mathcal{X}|\mathbf{e})p(\mathbf{e})$. The predicted likelihood of user-item interaction, denoted as $\hat{y}_{u,v}$, is derived by performing a dot product between the user and item representations, as follows: $\hat{y}_{u,v} = \mathbf{e}_{u}^\top \cdot \mathbf{e}_{v}$.

While state-of-the-art recommender systems operating within the ID-based collaborative filtering paradigm have exhibited remarkable performance, they face significant challenges when tasked with handling cold-start recommendation scenarios, especially in situations where data scarcity is prevalent. The primary obstacle that ID-based recommenders encounter in these situations stems from the lack of past interaction history for newly introduced items. This absence of accumulated user engagement data disrupts the optimization paradigm that these systems typically rely upon, thereby making it considerably more difficult to generate accurate and meaningful representations for these items. Consequently, ID-based recommenders may encounter considerable difficulties in effectively modeling and understanding the inherent characteristics and preferences associated with these cold-start items, leading to a notable decline in the overall performance, particularly in zero-shot scenarios where no prior interaction data is available for certain items.

\subsection{Text-empowered User/Item Representations}
To address the challenge of cold-start items in zero-shot recommendation scenarios, we propose a novel approach that leverages textual side features for generalized and adaptable user and item representation learning. Specifically, we seek to replace the traditional ID-based embeddings with the side information associated with the items, namely their text descriptions, represented as $\mathbf{F}\in\mathbb{R}^{|\mathcal{V}|\times d_t}$. To accomplish this, we utilize a project layer with a multi-layer perceptron $T_{raw}$ to map the raw textual features $\mathbf{f}\in\mathbb{R}^{d_t}$ into a lower-dimensional latent space $\mathbb{R}^d$. The resulting representation $\hat{\mathbf{f}} \in \mathbb{R}^{d}$ is then used as the initial item representation in our model.
\begin{equation}
    \hat{\mathbf{f}}_v  = T_{raw}(\mathbf{f}).
\end{equation}
This text-driven approach enables a seamless fusion of collaborative signals with textual semantics, empowering our recommender to capture user preferences with unparalleled accuracy. Using rich textual features as initial item representations, we optimize user ID embeddings based on observed interactions. This injects valuable collaborative cues into the user and item representations, making them adaptable to evolving preferences. This synergistic learning equips our model to deeply comprehend users' affinities for text-based items. Crucially, it also endows the system with zero-shot prediction capabilities for cold-start items lacking prior interaction data.

\noindent \textbf{LLM-enhanced User/Item Profiling}
To further empower user representations with the rich textual semantics provided by large language models (LLMs), we propose generating user profile information that can reflect their interaction preferences. Specifically, item profiles can be derived from the profiles of users who frequently interact with them. This approach proves valuable in capturing user preferences and facilitating accurate recommendations for cold-start items.

On the user side, the ID embedding $\mathbf{x}_u \in \mathbb{R}^{d}$ integrates with the user profile $\mathbf{p}_u \in \mathbb{R}^{d_t}$, allowing the system to leverage the user's ID-based embedding and their generated profile, which can capture more nuanced preferences. On the item side, the raw text $\mathbf{x}_v$ combines with the item profile $\mathbf{p}_v \in \mathbb{R}^{d_t}$, enabling the system to understand the item's characteristics and how they align with user preferences.
\begin{equation}
    \hat{\mathbf{f}}^{aug}_u = \Psi(\mathbf{x}_u\text{ | }T_{pro}(\mathbf{p}_u)),\quad \hat{\mathbf{f}}^{aug}_v = \Psi(\hat{\mathbf{f}}_v\text{ | }T_{pro}(\mathbf{p}_v)).
\end{equation}
Our framework employs a dual-MLP architecture for effective multi-modal fusion: an initial MLP encoder $\Psi(\cdot)$ consolidates heterogeneous features, followed by a profile transformation network $T_{pro}$ that projects the unified embeddings into the model's latent space. This hierarchical process generates enriched user and item representations $\hat{\mathbf{f}}^{aug}_u \in \mathbb{R}^{d}$ and $\hat{\mathbf{f}}^{aug}_v \in \mathbb{R}^{d}$, capturing comprehensive behavioral patterns. To further enhance representation quality, we leverage state-of-the-art LLMs as profile augmentation engines, generating supplementary semantic information that significantly boosts our recommender system's modeling capacity.

\begin{figure*}[t]
    \centering
    \includegraphics[width=1\linewidth]{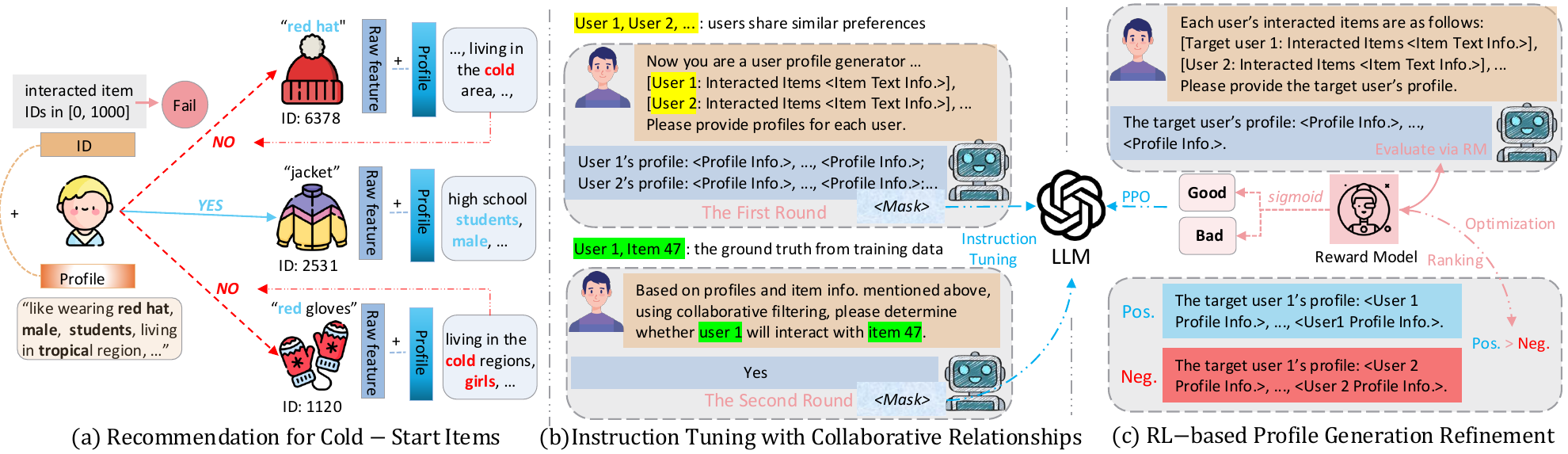}
    \caption{The Overall Framework of the Proposed \model.}
    \label{fig:figure_overall}
\end{figure*}

\subsection{Instruction Tuning with Collaborative Relationships}
\label{subsec:CF_llm}
Our \model\ framework seeks to align the collaborative relationships with the textual semantic representation space through an innovative recommendation instruction tuning paradigm that seamlessly integrates user collaborative relationships into the LLM-based profiling process. The framework employs a two-phase approach: first, it utilizes knowledge distillation and dialogue-based instruction tuning to preserve high-order collaborative similarities between users and items, generating high-quality and informative user profiles. Subsequently, it leverages these refined user profiles to generate semantically aligned item profiles, resulting in a cohesive representation space that effectively captures both individual characteristics and collaborative patterns.

\subsubsection{LLM Fine-Tuning with Collaborative Signals}
\label{sec:gpt_kd tuning}
Our approach leverages llama2-7b-chat as the base model, augmented with a novel collaborative instruction tuning framework that captures both user-item interactions and profile generation capabilities. Specifically, we design specialized prompt templates incorporating collaborative signals, and construct input-output pairs using ChatGPT-3.5 to capture diverse user-item relationship patterns. Here, the ChatGPT-3.5 is utilized as a powerful LLM with strong text summary capacity to generate text profile to characterize the interaction patters among suers and items, and the collaborative relationships among them, and result in textual instructions used for collaborative instruction tuning.

\subsubsection{Collaborative Instruction Tuning}
\label{sec:two_turn tuning}
In our recommendation instruction tuning paradigm, we propose to inject higher-order collaborative relational context through a novel two-turn collaborative instruction tuning paradigm. By incorporating collaborative signals that capture complex user-user and user-item relationships, this approach transcends the limitations of methods that rely solely on direct user-item interactions. Our framework serves a dual purpose: it empowers LLMs to generate richer and more nuanced user profiles by leveraging the collaborative network dynamics among users and items, while simultaneously addressing the fundamental challenge of insufficient direct supervision in profile generation tasks.

\noindent \textbf{Profile Generation with Two-turn Dialogue.} A fundamental challenge in LLM-based profile generation lies in the absence of ground truth data for direct evaluation, forcing practitioners to rely on indirect assessment through downstream recommendation performance. To address this limitation, we propose a novel two-turn collaborative instruction tuning paradigm that provides explicit supervision signals. Unlike conventional approaches that struggle with profile quality assessment, our framework leverages collaborative user relationships to guide LLMs in generating high-quality profiles through structured dialogue interactions.

$\bullet$ \textbf{First Turn - Collaborative Profile Generation:} The initial turn takes input query $\mathcal{Q}$ containing both the target user's historical item interactions and those of similar users identified from their collaborative neighborhood. Here, user similarities are measured by the encoded user embeddings from LightGCN model~\citep{he2020lightgcn}. The LLM generates output response $\mathcal{R}$ consisting of user profiles that capture collaborative patterns.
\begin{equation}
    \mathcal{Q}_{fir.} = Prompt(u_t, \{u_n\}, \mathcal{V}_t, \{\mathcal{V}_n\}),\quad \mathcal{R}_{fir.} = Prompt(u_t, \{u_n\}, \mathcal{P}_{t}, \{\mathcal{P}_{n}\}). 
\end{equation}

$\bullet$ \textbf{Second Turn - Supervised Interaction Prediction:} The second turn reformulates the instruction-tuning task as an interaction prediction problem, aiming to enhance profile generation by incorporating explicit user-item preference signals. Specifically, the input query $\mathcal{Q}$ prompts whether target user $u_t$ will interact with candidate item $v_t$, while the output response $\mathcal{R}$ corresponds to the ground truth interaction status (\textit{i.e.}, $yes$ or $no$) from the training dataset. To ensure balanced and informative training signals, we employ a systematic sampling strategy: For positive samples (50\%), we select $v^{+}$ from the user's interaction history $\mathcal{V}_t$ under the constraint that $v^{+}$ also appears in similar users' history ${\mathcal{V}_n}$, while removing it from $\mathcal{V}_t$ in first-turn instructions to prevent information leakage. For negative samples (50\%), we select $v^{-}$ from similar users' history ${\mathcal{V}_n}$ while ensuring no historical interaction exists between $u_t$ and $v^{-}$ in the training data. This balanced sampling approach maintains unbiased training objectives while preserving crucial collaborative signals for effective profile generation.

\begin{equation}
\begin{split}
    \mathcal{Q}_{sec.} = 
    \begin{cases}
        Prompt(u_t, v^{+}), &pos.\text{ }samp.\\
        Prompt(u_t, v^{-}), &neg.\text{ }samp.\\
    \end{cases}\quad \mathcal{R}_{sec.} = 
    \begin{cases}
        \text{Yes}, &\text{      } pos.\text{ }samp.\\
        \text{No}, &\text{      } neg.\text{ }samp.\\
    \end{cases}
\end{split}
\end{equation}

\noindent \textbf{Tuning Strategy.}
For multi-turn dialogue instruction-tuning, our objective is to utilize LLM-generated responses $\mathcal{R}$ for weight updates while excluding queries $\mathcal{Q}$. Traditional single-turn dialogue tuning, when applied to our paradigm, considers $\mathcal{Q}{fir.}$, $\mathcal{R}_{fir.}$, and $\mathcal{Q}_{sec.}$ as inputs, with only $\mathcal{R}_{sec.}$ being predicted. This approach limits weight updates to losses from $\mathcal{R}_{sec.}$ alone, failing to leverage the training data in multi-turn dialogues. Notably, $\mathcal{R}_{fir.}$ contains rich textual information as multiple user profiles, which guides the generation of $\mathcal{R}_{sec.}$ in subsequent turns. In contrast, $\mathcal{R}_{sec.}$ consists of simpler binary responses (e.g., $yes$ or $no$). Disregarding the valuable information in $\mathcal{R}_{fir.}$ and relying on $\mathcal{R}_{sec.}$ for fine-tuning would significantly limit the model's learning capacity.

To address these limitations, we introduce a two-turn dialogue tuning approach. Specifically, we concatenate dialogues from both turns and apply masking techniques to differentiate between $\mathcal{Q}$ and $\mathcal{R}$ components. During training, the model computes losses exclusively from $\mathcal{R}$-marked segments, enabling both $\mathcal{R}_{fir.}$ and $\mathcal{R}_{sec.}$ to contribute to parameter optimization. This unified training strategy maximizes information utilization while effectively capturing collaborative relationships.

\textbf{Inference Prompt.}
Following instruction-tuning, we design an inference prompt that integrates target user interactions $\mathcal{V}_t$ with neighbor histories ${\mathcal{V}_n}$. This unified prompt guides LLMs to generate collaborative-aware user profiles by leveraging inter-user preference patterns.
\begin{equation}
    \mathcal{Q}_{inf.} = Prompt(u_t, \{u_n\}, \mathcal{V}_t, \{\mathcal{V}_n\}).
\end{equation}

\subsection{Refining Profile Generation through Reinforcement Learning}
\label{subsec:RL_llm}
We propose a reinforcement learning framework to enhance LLM-generated user profiles, focusing on improving accuracy and personalization. Our approach addresses two fundamental challenges: \textbf{i) Prompt-Training Discrepancy}—the inherent mismatch between inference (single-profile generation) and fine-tuning (multi-profile generation) phases that introduces potential inconsistencies in profile generation, and \textbf{ii) Personalization-Collaboration Trade-off}—while leveraging collaborative information enhances overall performance, it risks diluting individual user characteristics, analogous to the over-smoothing phenomenon observed in our collaborative instruction-tuning paradigm.

To address these challenges, we develop a reinforcement learning-based fine-tuning paradigm inspired by RLHF (Reinforcement Learning from Human Feedback)\citep{stiennon2020learning}. Our approach consists of two key components: a reward model that evaluates LLM-generated profile quality, and an optimization procedure using Proximal Policy Optimization (PPO)\citep{schulman2017proximal} that refines the LLM based on these reward signals. Through this iterative process, we progressively enhance the LLM's ability to generate more accurate and personalized user profiles.

\noindent \textbf{Reward Model.} We design a reward model to evaluate the quality of LLM-generated outputs from a human perspective. Specifically, given an input pair [$\mathcal{Q}$, $\mathcal{R}$], the model produces a scalar value indicating the response quality. The reward model is optimized using the loss function $\mathcal{L}_{rm}$:
\begin{equation}
    \mathcal{L}_{rm} = -\sum_{i=0}^{\mathcal{I}}\mathbb{E}_{(\mathcal{Q}_i,\mathcal{R}_i^{+}, \mathcal{R}_i^{-})\sim D}[\text{log}(\sigma(r_{\theta}(\mathcal{Q}_i, \mathcal{R}_i^{+})-r_{\theta}(\mathcal{Q}_i, \mathcal{R}_i^{-})))],
\end{equation}
\noindent where $r_{\theta}(\cdot)$ denotes the reward model, $\sigma(\cdot)$ represents the sigmoid function, and $\mathcal{R}_i^{+}$/$\mathcal{R}_i^{-}$ indicate true/false responses respectively. For our profiling task, we maintain a consistent query $\mathcal{Q}_{inf.}$ while carefully constructing both positive responses $\mathcal{R}^{+}$ and negative responses $\mathcal{R}^{-}$. We obtain $\mathcal{R}^{+}$ through ChatGPT and create $\mathcal{R}^{-}$ through two strategies:
\begin{itemize}[leftmargin=*]
\item \textbf{Diverse Negative Sampling:} Multiple prompt templates generate varied low-quality responses, enabling the reward model to identify suboptimal outputs post instruction-tuning.
\item \textbf{Profile Substitution:} Strategic replacement with similar users' profiles helps the model discern subtle differences and prioritize personalized characteristics.
\end{itemize}

\textbf{Proximal Policy Optimization.}
In our reinforcement learning framework, we treat the LLM $\mathcal{M}$ as the policy to be optimized, with the reward model approximating the true reward function. The core optimization objective is formulated as:
\begin{equation}
    \mathop{\text{argmax}}\limits _{\mathcal{M}}\mathbb{E}_{x_i\sim \mathcal{D}, y_i\sim \mathcal{M}}[R(y_i | x_i)].
\end{equation}
To optimize $\mathcal{M}$ iteratively, we sample queries $\mathcal{Q}i$ from the dataset $\mathcal{D}$ and generate corresponding responses $\mathcal{R}i$ using $\mathcal{M}$. We employ the Proximal Policy Optimization (PPO) algorithm and its associated loss function for optimization. Following~\cite{schulman2017proximal}, we enhance the reward function with a KL divergence penalty term between the original LLM $\mathcal{M}_{0}$ and the optimized LLM $\mathcal{M}_{\theta}$. This constraint effectively mitigates reward hacking—a phenomenon where models achieve high reward scores but poor human evaluation results. The final reward function $R(\cdot)$ for a given query-response pair $(\mathcal{Q}_i, \mathcal{R}_i)$ is defined as:
\begin{equation}
\begin{split}
    R(\mathcal{R}_i | \mathcal{Q}_i) &= \hat{r}(\mathcal{R}_i | \mathcal{Q}_i) - \beta D_{KL}(\mathcal{M}_{\theta}(\mathcal{Q}_i) || \mathcal{M}_{0}(\mathcal{Q}_i))
\end{split}
\end{equation}
The comprehensive details of our instruction design across all fine-tuning stages, together with our methodology for constructing positive and negative training samples for the reinforcement learning-based reward model, are thoroughly documented in Appendix~\ref{app:ins}.

\section{Evaluation}
\label{sec:eval}
In this section, we verify the effectiveness of \model by answering the following several questions:
\begin{itemize}[leftmargin=*]
\item \textbf{RQ1}: How does our proposed \model enhance the performance of existing recommender systems, particularly in item cold-start scenarios?

\item \textbf{RQ2}:  What contributions do the instruction-tuning techniques and reinforcement learning enhancements make to overall recommendation performance?

\item \textbf{RQ3}: How effective is our LLM-empowered user/item profiling system as an embedding function?

\item \textbf{RQ4}: How does our method perform in terms of efficiency?

\item \textbf{RQ5}: What advantages does our method have compared to existing LLM-enhanced recommenders? 

\item \textbf{RQ6}: How does the reinforcement learning-based profile generation refinement enhance the performance of our LLM-empowered profiling system?

\end{itemize}

\subsection{Experimental Settings}
\label{sec:exp_setting}
To evaluate the effectiveness of our proposed method, we conduct extensive experiments using two public datasets: \textbf{MIND}\footnote{https://msnews.github.io}~\citep{wu2020mind} and \textbf{Netflix}\footnote{https://www.kaggle.com/datasets/netflix-inc/netflix-prize-data}, along with a large-scale dataset derived from real-world industrial data (referred to as Industrial for anonymity).

We assess the accuracy of the top-\textit{K} recommendation results using two widely adopted metrics: Recall@K (R@K) and NDCG (N@K), with $K$ set to 20 by default. To reduce bias, we employ an all-rank evaluation strategy, where positive items in the test set are ranked alongside all non-interacted items for each user. The final metric is reported as the average score across all users in the test set.

We evaluate the effectiveness of our \model approach by integrating it with state-of-the-art recommender systems, allowing us to assess performance improvements in a model-agnostic manner compared to baseline models. The selected CF recommenders include non-graph methods such as BiasMF~\citep{koren2009matrix} and NCF~\citep{he2017neural}, the GNN-enhanced method LightGCN~\citep{he2020lightgcn}, and graph contrastive learning approaches SGL~\citep{wu2021self} and SimGCL~\citep{yu2022graph}. Details regarding the datasets and baseline methods are provided in Appendices~\ref{app:dataset} and~\ref{app:baseline}.

\subsection{Performance Comparison (RQ1)}
\label{sec:rq1}

\begin{table*}[!htbp]
    \centering
    \small
    \caption{Performance comparison on MIND, Netflix and Industrial data in terms of \textit{Recall} and \textit{NDCG}. The superscript * indicates the improvement is statistically significant where the p-value $< 0.05$.}
    \setlength{\tabcolsep}{0.3mm}
    \resizebox{\linewidth}{!}{
        \begin{tabular}{c|c|l l l l|l l l l|l l l l}
            \hline
            \multicolumn{2}{c|}{Dataset} & \multicolumn{4}{|c|}{MIND} & \multicolumn{4}{|c|}{Netflix} & \multicolumn{4}{|c}{Industrial} \\
            \hline
            Backbone & Variants & R@20 & R@40 & N@20 & N@40 & R@20 & R@40 & N@20 & N@40 & R@20 & R@40 & N@20 & N@40 \\
            \hline
            \multicolumn{14}{c}{Full-Shot Setting} \\
            \hline
            \multirow{3}{*}{BiasMF} & Base & $0.0683$ & $0.1039$ & $\textbf{0.0311}$ & $0.0399$ & $0.0449$ & $0.0790$ & $0.1451$ & $0.1375$ & $0.0078$ & $0.0143$ & $0.0046$ & $0.0066$ \\
            & Augment. & $\textbf{0.0719}^{*}$ & $\textbf{0.1353}^{*}$ & 0.0272 & $\textbf{0.0411}^{*}$ & $\textbf{0.0531}^{*}$ & $\textbf{0.0868}^{*}$ & $\textbf{0.1761}^{*}$ & $\textbf{0.1630}^{*}$ & $\textbf{0.0121}^{*}$ & $\textbf{0.0198}^{*}$ & $\textbf{0.0074}^{*}$ & $\textbf{0.0097}^{*}$ \\
            & \textbf{Improve.} & $5.27\%\uparrow$ & $30.22\%\uparrow$ & $12.54\%\downarrow$ & $3.01\%\uparrow$ & $18.26\%\uparrow$ & $9.87\%\uparrow$ & $21.36\%\uparrow$ & $18.55\%\uparrow$ & $55.13\%\uparrow$ & $38.46\%\uparrow$ & $60.87\%\uparrow$ & $46.97\%\uparrow$ \\
            \hline
            \multirow{3}{*}{NCF} & Base & $0.0713$ & $0.0985$ & $\textbf{0.0325}$ & $\textbf{0.0445}$ & $0.0581$ & $0.0936$ & $0.1848$ & $0.1721$ & $0.0102$ & $0.0076$ & $0.0188$ & $0.0091$ \\
            & Augment. & $\textbf{0.0760}^{*}$ & $\textbf{0.1233}^{*}$ & $0.0288$ & $0.0414$ & $\textbf{0.0591}^{*}$ & $\textbf{0.0968}^{*}$ & $\textbf{0.1903}^{*}$ & $\textbf{0.1785}^{*}$ & $\textbf{0.0133}^{*}$ & $\textbf{0.0087}^{*}$ & $\textbf{0.0206}^{*}$ & $\textbf{0.0108}*{*}$ \\
            & \textbf{Improve.} & $6.59\%\uparrow$ & $25.18\%\uparrow$ & $11.38\%\downarrow$ & $6.97\%\downarrow$ & $1.72\%\uparrow$ & $3.42\%\uparrow$ & $2.98\%\uparrow$ & $3.72\%\uparrow$ & $30.39\%\uparrow$ & $14.47\%\uparrow$ & $9.57\%\uparrow$ & $18.68\%\uparrow$ \\
            \hline
            \multirow{3}{*}{LightGCN} & Base & $0.0389$ & $0.0702$ & $0.0150$ & $0.0219$ & $0.0467$ & $0.0815$ & $0.1488$ & $0.1424$ & $0.0096$ & $0.0162$ & $0.0059$ & $0.0076$ \\
            & Augment. & $\textbf{0.0788}^{*}$ & $\textbf{0.0983}^{*}$ & $\textbf{0.0337}^{*}$ & $\textbf{0.0384}^{*}$ & $\textbf{0.0652}^{*}$ & $\textbf{0.1026}^{*}$ & $\textbf{0.1703}^{*}$ & $\textbf{0.1606}^{*}$ & $\textbf{0.0143}^{*}$ & $\textbf{0.0225}^{*}$ & $\textbf{0.0087}^{*}$ & $\textbf{0.0107}^{*}$ \\
            & \textbf{Improve.} & $102.57\%\uparrow$ & $40.03\%\uparrow$ & $124.67\%\uparrow$ & $75.34\%\uparrow$ & $39.61\%\uparrow$ & $25.89\%\uparrow$ & $14.45\%\uparrow$ & $12.78\%\uparrow$ & $48.96\%\uparrow$ & $38.89\%\uparrow$ & $47.46\%\uparrow$ & $40.79\%\uparrow$ \\
            \hline
            \multirow{3}{*}{SGL} & Base & $0.0345$ & $0.0708$ & $0.0127$ & $0.0210$ & $0.0277$ & $0.0416$ & $0.0855$ & $0.0762$ & $0.0078$ & $0.0138$ & $0.0050$ & $0.0068$ \\
            & Augment. & $\textbf{0.0732}^{*}$ & $\textbf{0.0967}^{*}$ & $\textbf{0.0367}^{*}$ & $\textbf{0.0421}^{*}$ & $\textbf{0.0788}^{*}$ & $\textbf{0.1204}^{*}$ & $\textbf{0.1958}^{*}$ & $\textbf{0.1831}^{*}$ & $\textbf{0.0133}^{*}$ & $\textbf{0.0221}^{*}$ & $\textbf{0.0080}^{*}$ & $\textbf{0.0106}^{*}$ \\
            & \textbf{Improve.} & $112.17\%\uparrow$ & $36.58\%\uparrow$ & $188.98\%\uparrow$ & $100.48\%\uparrow$ & $184.48\%\uparrow$ & $189.42\%\uparrow$ & $129.01\%\uparrow$ & $140.29\%\uparrow$ & $70.51\%\uparrow$ & $60.14\%\uparrow$ & $60\%\uparrow$ & $55.88\%\uparrow$ \\
            \hline
            \multirow{3}{*}{SimGCL} & Base & $0.0421$ & $0.0636$ & $0.0155$ & $0.0212$ & $0.0231$ & $0.0441$ & $0.0810$ & $0.0825$ & $0.0042$ & $0.0078$ & $0.0026$ & $0.0037$ \\
            & Augment. & $\textbf{0.0576}^{*}$ & $\textbf{0.0908}^{*}$ & $\textbf{0.0232}^{*}$ & $\textbf{0.0329}^{*}$ & $\textbf{0.0567}^{*}$ & $\textbf{0.0908}^{*}$ & $\textbf{0.1782}^{*}$ & $\textbf{0.1673}^{*}$ & $\textbf{0.0128}^{*}$ & $\textbf{0.0205}^{*}$ & $\textbf{0.0080}^{*}$ & $\textbf{0.0099}^{*}$ \\
            & \textbf{Improve.} & $36.82\%\uparrow$ & $42.77\%\uparrow$ & $49.68\%\uparrow$ & $55.19\%\uparrow$ & $145.45\%\uparrow$ & $105.90\%\uparrow$ & $120.00\%\uparrow$ & $102.79\%\uparrow$ & $204.76\%\uparrow$ & $162.82\%\uparrow$ & $207.69\%\uparrow$ & $167.57\%\uparrow$ \\
            \hline
            \multicolumn{14}{c}{Zero-Shot Setting} \\
            \hline
            \multirow{3}{*}{BiasMF} & Base & $0.0096$ & $0.0165$ & $0.0031$ & $0.0041$ & $0.0311$ & $0.0769$ & $0.0167$ & $0.0292$ & $0.0038$ & $0.0068$ & $0.0020$ & $0.0029$ \\
            & Augment. & $\textbf{0.0246}^{*}$ & $\textbf{0.0373}^{*}$ & $\textbf{0.0107}^{*}$ & $\textbf{0.0135}^{*}$ & $\textbf{0.1381}^{*}$ & $\textbf{0.1490}^{*}$ & $\textbf{0.0828}^{*}$ & $\textbf{0.0584}^{*}$ & $\textbf{0.0056}^{*}$ & $\textbf{0.0103}^{*}$ & $\textbf{0.0026}^{*}$ & $\textbf{0.0040}^{*}$ \\
            & \textbf{Improve.} & $156.25\%\uparrow$ & $126.06\%\uparrow$ & $245.16\%\uparrow$ & $229.27\%\uparrow$ & $344.05\%\uparrow$ & $93.76\%\uparrow$ & $395.81\%\uparrow$ & $100.00\%\uparrow$ & $47.37\%\uparrow$ & $51.47\%\uparrow$ & $30.00\%\uparrow$ & $37.93\%\uparrow$ \\
            \hline
            \multirow{3}{*}{NCF} & Base & $0.0301$ & $0.0383$ & $0.0080$ & $0.0097$ & $0.0480$ & $0.1158$ & $0.0196$ & $0.0384$ & $0.0044$ & $0.0022$ & $0.0056$ & $0.0026$ \\
            & Augment. & $\textbf{0.0424}^{*}$ & $\textbf{0.0469}^{*}$ & $\textbf{0.0112}^{*}$ & $\textbf{0.0122}^{*}$ & $\textbf{0.1700}^{*}$ & $\textbf{0.1774}^{*}$ & $\textbf{0.0984}^{*}$ & $\textbf{0.0974}^{*}$ & $\textbf{0.0051}^{*}$ & $\textbf{0.0031}^{*}$ & $\textbf{0.0088}^{*}$ & $\textbf{0.0041}^{*}$ \\
            & \textbf{Improve.} & $40.86\%\uparrow$ & $22.45\%\uparrow$ & $40.00\%\uparrow$ & $25.77\%\uparrow$ & $254.17\%\uparrow$ & $53.20\%\uparrow$ & $402.04\%\uparrow$ & $153.65\%\uparrow$ & $15.91\%\uparrow$ & $40.91\%\uparrow$ & $57.14\%\uparrow$ & $57.69\%\uparrow$ \\
            \hline
            \multirow{3}{*}{LightGCN} & Base & $0.0138$ & $0.0292$ & $0.0046$ & $0.0078$ & $0.0974$ & $0.1256$ & $0.0446$ & $0.0415$ & $0.0092$ & $0.0160$ & $0.0051$ & $0.0070$ \\
            & Augment. & $\textbf{0.0196}^{*}$ & $\textbf{0.0389}^{*}$ & $\textbf{0.0064}^{*}$ & $\textbf{0.0086}^{*}$ & $\textbf{0.1371}^{*}$ & $\textbf{0.1453}^{*}$ & $\textbf{0.0697}^{*}$ & $\textbf{0.0459}^{*}$ & $\textbf{0.0133}^{*}$ & $\textbf{0.0188}^{*}$ & $\textbf{0.0090}^{*}$ & $\textbf{0.0106}^{*}$ \\
            & \textbf{Improve.} & $42.03\%\uparrow$  & $33.22\%\uparrow$  & $39.13\%\uparrow$ & $10.26\%\uparrow$  & $40.76\%\uparrow$  & $15.68\%\uparrow$  & $56.28\%\uparrow$ & $10.60\%\uparrow$  & $44.57\%\uparrow$  & $17.50\%\uparrow$  & $76.47\%\uparrow$ & $51.43\%\uparrow$  \\
            \hline
            \multirow{3}{*}{SGL} & Base & $0.0162$ & $0.0264$ & $0.0062$ & $0.0074$ & $0.0385$ & $0.1441$ & $0.0274$ & $0.0579$ & $0.0065$ & $0.0114$ & $0.0036$ & $0.0050$ \\
            & Augment. & $\textbf{0.0254}^{*}$ & $\textbf{0.0450}^{*}$ & $\textbf{0.0089}^{*}$ & $\textbf{0.0107}^{*}$ & $\textbf{0.1126}^{*}$ & $\textbf{0.1756}^{*}$ & $\textbf{0.0384}^{*}$ & $\textbf{0.1066}^{*}$ & $\textbf{0.0111}^{*}$ & $\textbf{0.0176}^{*}$ & $\textbf{0.0066}^{*}$ & $\textbf{0.0084}^{*}$ \\
            & \textbf{Improve.} & $56.79\%\uparrow$ & $70.45\%\uparrow$ & $43.55\%\uparrow$ & $44.59\%\uparrow$ & $92.47\%\uparrow$ & $21.86\%\uparrow$ & $40.15\%\uparrow$ & $84.11\%\uparrow$ & $70.77\%\uparrow$ & $54.39\%\uparrow$ & $83.33\%\uparrow$ & $68.00\%\uparrow$ \\
            \hline
            \multirow{3}{*}{SimGCL} & Base & $0.0164$ & $0.0300$ & $0.0055$ & $0.0084$ & $0.0793$ & $0.1259$ & $0.0336$ & $0.0460$ & $0.0078$ & $\textbf{0.0140}$ & $0.0042$ & $0.0059$ \\
            & Augment. & $\textbf{0.0312}^{*}$ & $\textbf{0.0388}^{*}$ & $\textbf{0.0098}^{*}$ & $\textbf{0.0115}^{*}$ & $\textbf{0.1508}^{*}$ & $\textbf{0.1895}^{*}$ & $\textbf{0.1550}^{*}$ & $\textbf{0.1647}^{*}$ & $\textbf{0.0084}^{*}$ & $0.0137$ & $\textbf{0.0044}^{*}$ & $0.0059$ \\
            & \textbf{Improve.} & $90.24\%\uparrow$ & $29.33\%\uparrow$ & $78.18\%\uparrow$ & $36.90\%\uparrow$ & $90.16\%\uparrow$ & $50.52\%\uparrow$ & $361.31\%\uparrow$ & $258.04\%\uparrow$ & $7.69\%\uparrow$ & $2.14\%\downarrow$ & $4.76\%\uparrow$ & \rule{0.7cm}{0.4pt} \\
            \hline
        \end{tabular}
    }
    \label{tab:exp_overall}
\end{table*}

To demonstrate the effectiveness of our \model in enhancing performance, particularly in cold-start scenarios, we apply it to five common collaborative filtering methods. The "full-shot" setting corresponds to the complete dataset, while the "zero-shot" setting refers to the pure cold-start condition. The \textbf{\textit{Base}} variant applies the cold-start recommendation paradigm to the baseline recommenders without any profiling enhancement via LLMs, whereas the \textbf{\textit{Augment}} variant integrates \model into the base recommenders. Detailed settings and implementation information are provided in Appendices~\ref{app:overall} and~\ref{app:imp}. The evaluation results in Table~\ref{tab:exp_overall} reveal several interesting observations.

\noindent \textbf{(i) Performance Improvement in Integrated Recommenders}. We consistently find that integrating \model with backbone recommenders leads to enhanced performance compared to the base variant, which relies on raw external item features and ID-based user embeddings in both \textbf{supervised} and \textbf{zero-shot} settings. This provides compelling evidence for the effectiveness of \model. We attribute these improvements to two key factors: \emph{First}, for supervised recommendation scenarios, \model leverages instruction-tuned LLMs to generate accurate user and item profiles as auxiliary information, effectively enhancing the semantic representation of user preference. \emph{Second}, our tuning paradigm guides the LLMs in capturing user collaborative relationships, allowing for the generation of high-quality, personalized profiles that demonstrate strong generalization in zero-shot scenarios.

\noindent \textbf{(ii) Outstanding Performance in Cold-Start Scenarios.} This improvement arises from our innovative modifications to the ID-embedding paradigm employed in current recommenders. By incorporating external features specifically designed to address the challenges of interaction data scarcity, we have significantly enhanced the effectiveness of these systems. Remarkably, we observe substantial performance improvements even in the relatively sparser MIND and Industrial datasets, where data limitations traditionally pose significant hurdles. By leveraging our \model for user and item profiling, we significantly enhance the generalization capabilities of existing recommenders.

\noindent \textbf{(iii) Practicality and Scalability for Real-World Deployment.} The results from the Industrial dataset demonstrate that \model consistently enhances the performance of recommenders in large-scale, highly sparse real-world scenarios. Furthermore, our user and item profile generation methods can be efficiently executed as an offline profiling system to support online applications, making them highly practical for real-world recommendations. To facilitate online recommendation systems, user and item profiles can be updated at regular intervals, such as daily or weekly. The performance improvements observed across various backbone models indicate that \model can easily adapt to a range of business models, significantly enhancing their overall effectiveness.

\subsection{Ablation Study (RQ2)}

\begin{figure}[h]
    \centering
    \subfigure[MIND data]{
    \label{fig:figure_exp_llm_ab_mind}
    \includegraphics[width=0.24 \columnwidth]{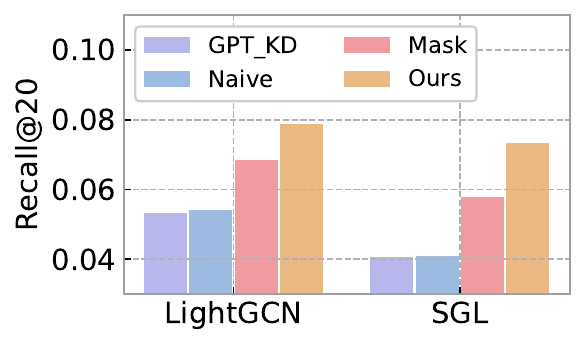}
    \includegraphics[width=0.24 \columnwidth]{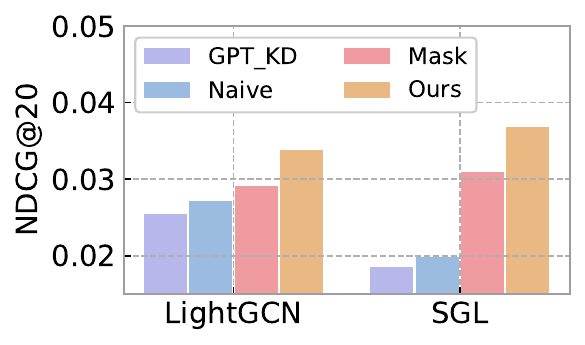}}
    \subfigure[Netflix data]{
    \label{fig:figure_exp_llm_ab_netflix}
    \includegraphics[width=0.24 \columnwidth]{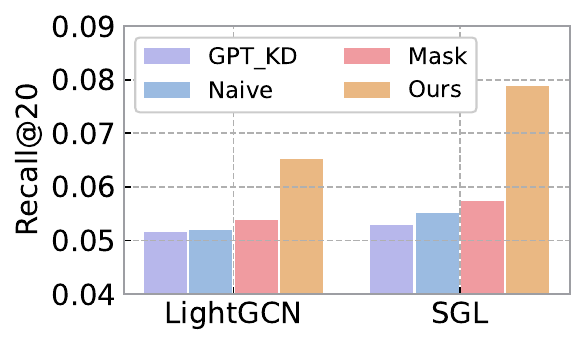}
    \includegraphics[width=0.24 \columnwidth]{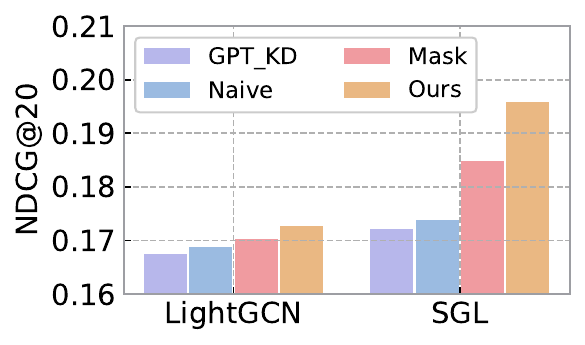}}
    \caption{Ablation study on the LLM tuning techniques in the \model\ framework.}
    \vspace{-0.1in}
    \label{fig:figure_exp_llm_ab}
\end{figure}

We conducted extensive experiments to validate the effectiveness of our proposed instruction tuning techniques by customizing three variants of \model: \textit{GPT\_KD}, \textit{Naive}, and \textit{Mask}. Detailed descriptions of these variants can be found in Appendix~\ref{app:ablation}. The results of our experiments are illustrated in Figure~\ref{fig:figure_exp_llm_ab}, allowing us to draw the following conclusions:

\noindent \textbf{(i) Advantage of Collaborative Instruction Tuning.} 
The results in Figure~\ref{fig:figure_exp_llm_ab} show that using instruction tuning to capture collaborative relationships among users and items, along with the masking tuning strategy (\textit{Mask}), significantly enhances performance compared to \textit{GPT\_KD}. This improvement suggests that our tuning solution generates more precise, high-quality profiles by leveraging collaborative information effectively. In contrast, profiling based solely on user interaction history has limitations, as it lacks the guidance from collaborative insights. Consequently, this approach often results in less accurate profiles that may include noisy interaction records.

\noindent \textbf{(ii) Effectiveness of the Masking-Based Tuning Strategy.} Although the \textit{Naive} variant also employs a two-round dialogue-based instruction tuning technique similar to the \textit{Mask} variant, its improvement over the \textit{GPT\_KD} variant is limited. This underscores the advantages of the masking-based tuning strategy, which effectively utilizes responses from the two-round dialogue to update the weights of the LLM and guide its learning of collaborative relationships between users.

\noindent \textbf{(iii) Benefits of Reinforcement Learning-Based Profile Generation Refinement.} The results indicate that the \textit{Mask} variant performs significantly worse than \model. This finding suggests that the proposed reinforcement learning (RL)-based profile generation refinement technique effectively addresses the noise issues and over-smoothing problems associated with the collaborative instruction-tuning paradigm. As a result, it enables the LLM to generate more accurate and personalized profiles.

\subsection{Effectiveness of LLM-empowered Profiling System in \model (RQ3)}
\begin{wraptable}{hr}{0.5\columnwidth}
    \centering
    \caption{Performance \textit{w.r.t.} various aug. variants.}
    \resizebox{\linewidth}{!}{
        \begin{tabular}{c|c|c c|c c}
            \hline
            \multicolumn{2}{c|}{Dataset} & \multicolumn{2}{|c|}{MIND} & \multicolumn{2}{|c}{Netflix} \\
            \hline
            Backbone & Variants & R@20 & N@20 & R@20 & N@20 \\
            \hline
            \multirow{4}{*}{LightGCN} & Base & 0.0389 & 0.0150 & 0.0467 & 0.1488 \\
            & w/o User Aug. & 0.0302 & 0.0123 & 0.0384 & 0.1213 \\
            & w/o Item Aug. & 0.0719 & 0.0287 & 0.0505 & 0.1621 \\
            & \model & \textbf{0.0788} & \textbf{0.0337} & \textbf{0.0652} & \textbf{0.1703} \\
            \hline
            \multirow{4}{*}{SGL} & Base & 0.0345 & 0.0127 & 0.0277 & 0.0855 \\
            & w/o User Aug. & 0.0253 & 0.0093 & 0.0173 & 0.0578 \\
            & w/o Item Aug. & 0.0719 & 0.0289 & 0.0502 & 0.1546 \\
            & \model & \textbf{0.0732} & \textbf{0.0367} & \textbf{0.0788} & \textbf{0.1958} \\
            \hline
        \end{tabular}
    }
    \vspace{-0.1in}
    \label{tab:exp_rec_ab} 
\end{wraptable}
To investigate the impact of our LLM-empowered profiling system on user and item feature enhancements, we developed two variants of \model: one that excludes user feature enhancement (\textit{i.e.}, w/o User Aug.) and another that excludes item feature enhancement (\textit{i.e.}, w/o Item Aug.). The experiments were conducted on the MIND and Netflix datasets using the full-shot setting, with LightGCN and SGL as the backbone models. The evaluation results are presented in Table~\ref{tab:exp_rec_ab}, allowing us to draw the following conclusions.


\noindent \textbf{(i) User-Side Feature Enhancement.} The exclusion of user-side feature enhancements (\textit{i.e.}, w/o User Aug.) results in a significant decline in performance across both evaluated datasets and backbone models. This underscores the critical role of our \model as the profiling system for improving performance. Relying solely on the original ID embedding for the user side is insufficient for effectively capturing and modeling user preferences. We attribute this outcome to both the effective extraction of text features and the successful integration of graph and textual information.


\noindent \textbf{(ii) Item-Side Feature Enhancement.} The exclusion of item-side feature enhancements (\textit{i.e.}, w/o Item Aug.) also leads to a noticeable decline in the recommender's performance. Interestingly, when item-side feature enhancements are retained without incorporating any user-side feature enhancements (\textit{i.e.}, w/o User Aug.), the performance can drop even below that of the Base variant. This discrepancy can be attributed to the interplay between raw and enhanced features on the item side, which creates a complex dynamic. Relying solely on ID embedding for the user side proves inadequate for effectively modeling user preferences.

\subsection{Training Efficiency Analysis of \model (RQ4)}
\label{sec:train_eff}
\begin{wraptable}{r}{0.4\columnwidth}
    \centering
    \caption{Training efficiency \textit{w.r.t.} integration with various recommenders.}
    \vspace{-0.05in}
    \resizebox{\linewidth}{!}{
    \footnotesize
    \begin{tabular}{c|c|c|c|c}
        \hline
         Dataset & Recommender & Base & \model & Cost \\
         \hline
         \hline
         \multirow{5}{*}{MIND} & BiasMF & 0.72s &0.85s &$+18.06\%$ \\
         & NCF & 0.76s& 0.85s&$+11.84\%$ \\
         & LightGCN &0.79s & 0.86s& $+8.86\%$\\
         & SGL &1.93s &2.01s & $+4.15\%$\\
         & SimGCL &2.63s & 2.69s& $+2.28\%$\\
         \hline
         \multirow{5}{*}{Netflix} & BiasMF & 14.38s &16.42s & $+14.19\%$\\
         & NCF &15.02s &17.17s &$+14.31\%$ \\
         & LightGCN &20.47s &20.95s &$+2.34\%$ \\
         & SGL &64.98s &65.08s & $+0.15\%$\\
         & SimGCL &44.02s &44.61s & $+1.34\%$\\
         \hline
         \multirow{5}{*}{Industrial} & BiasMF &7.07s  &8.85s & $+25.18\%$\\
         & NCF &7.58s &8.45s & $+11.48\%$\\
         & LightGCN & 9.33s&10.25s & $+9.86\%$\\
         & SGL & 32.34s& 32.87s& $+1.64\%$\\
         & SimGCL & 85.41s&86.52s & $+1.30\%$\\
         \hline
    \end{tabular}
    }
    \label{tab:exp_eff}
    \vspace{-0.15in}
\end{wraptable}
To evaluate the efficiency of our \model\ approach, we conduct both a theoretical complexity analysis and an empirical running time test. \textbf{Theoretical Analysis}: The time complexity of the MLP used to transfer textual features $\mathbf{f} \in \mathbb{R}^{d_t}$ of items into the model's latent space $\mathbb{R}^{d}$ is $\mathcal{O}(N \times (d_t \times d + d\times d))$, where $N$ represents the number of nodes, and $d_t$ and $d$ denote the dimensionalities of the original text features and the latent space, respectively. \textbf{Empirical Evaluation}: We present the per-epoch training time in Table~\ref{tab:exp_eff}. The evaluation was conducted on a server equipped with NVIDIA A100 GPUs (40 GB memory). The results indicate that for larger models (e.g., GNN-based methods), our \model requires relatively little additional time, often falling below 10\%. In denser datasets like Netflix, this additional time can be reduced to under 5\%. Even for smaller recommenders, the maximum additional time is approximately 25\%. Given the substantial improvements in recommendation provided by our method, the incurred costs are considered acceptable.

\subsection{Comparison with existing LLM-enhanced methods(RQ5)}
\label{sec:compare_LLMRec}
\begin{wrapfigure}{hr}{0.5\columnwidth}
    \centering
    \vspace{-0.25in}
    \subfigure[MIND data]{
    \label{fig:figure_exp_llmrec_mind}
    \includegraphics[width=0.25 \columnwidth]{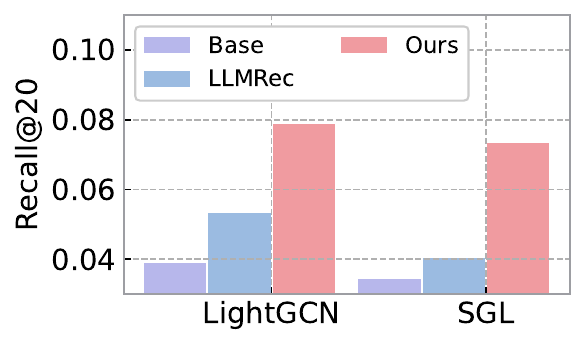}
    \includegraphics[width=0.25 \columnwidth]{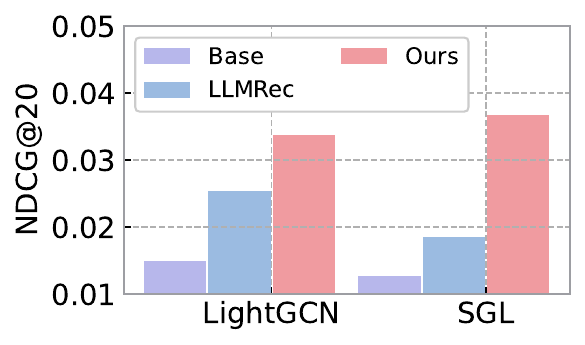}}
    \subfigure[Netflix data]{
    \label{fig:figure_exp_llmrec_netflix}
    \includegraphics[width=0.25 \columnwidth]{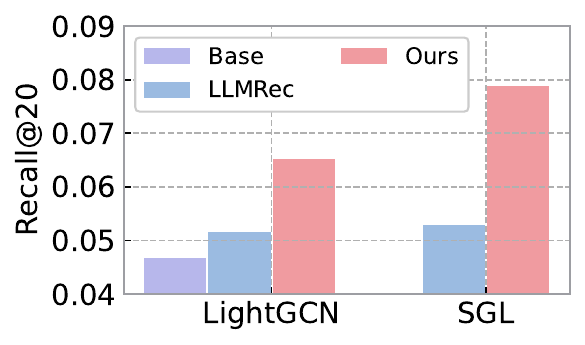}
    \includegraphics[width=0.25 \columnwidth]{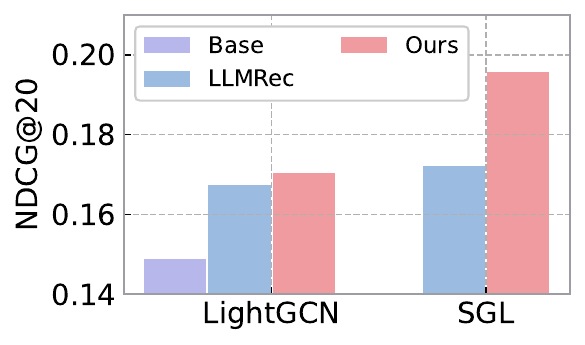}}
    \vspace{-0.1in}
    \caption{Performance Comparison with LLMRec.}
    \label{fig:exp_comp_llmrec}
    \vspace{-0.25in}
\end{wrapfigure}

We further compare \model with the existing work LLMRec~\citep{wei2024llmrec}, which also enhances recommendation systems using LLMs, to highlight the superiority of our proposed instruction-tuning technique. The experimental results are presented in Figure~\ref{fig:exp_comp_llmrec}. Specifically, LLMRec generates profiles for items and users by directly calling the LLM's API without fine-tuning for the profile generation task. This approach fails to effectively leverage the collaborative relationships among users. As a result, \model demonstrates significant performance advantages across two public datasets, leading to notable improvements in the performance of the base models.

\subsection{Case Study(RQ6)}

\begin{wrapfigure}{r}{0.5\columnwidth}
    \vspace{-0.1in}
    \centering
    \includegraphics[width=0.5\columnwidth]{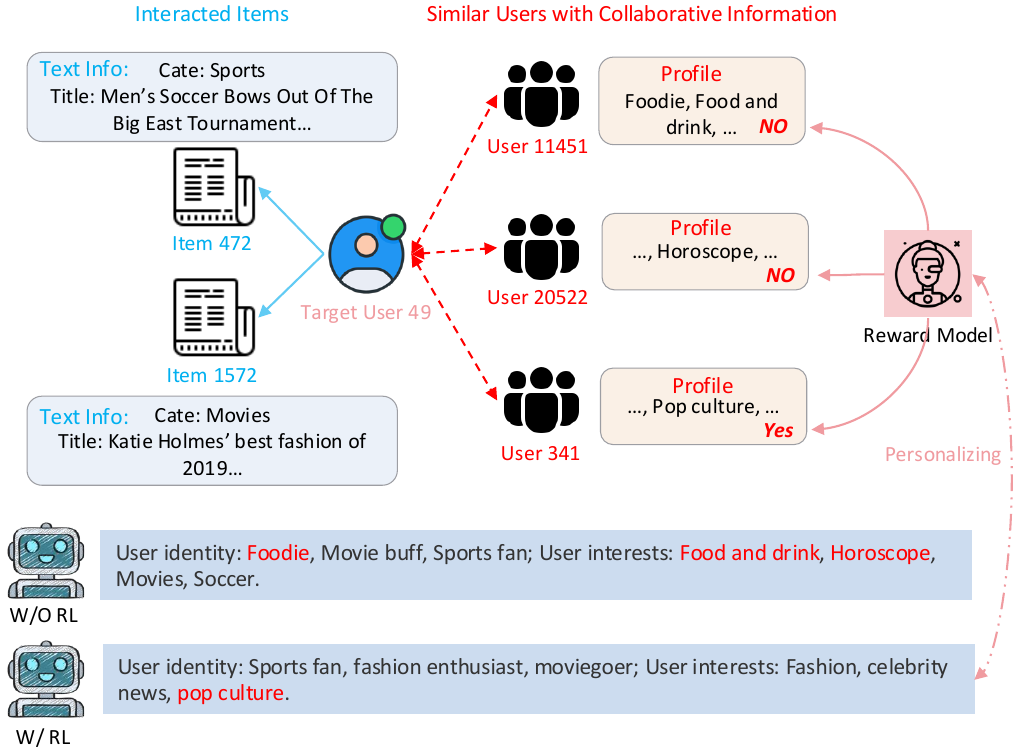}
    \vspace{-0.2in}
    \caption{Generated profiles w/ and w/o RL.}
    \vspace{-0.25in}
    \label{fig:figure_Case}
\end{wrapfigure}

To intuitively explore the contribution of reinforcement learning to the personalization of generated profiles, we conducted a case study using the MIND dataset. In this study, as shown in Figure~\ref{fig:figure_Case}, the target user for whom the profile is being generated is \textit{User 49}. This user has interacted with two items: \textit{Item 472} and \textit{Item 1572}. Additionally, we identified three similar users who provide collaborative information: \textit{User 11451}, \textit{User 20522}, and \textit{User 341}.

The user profile generated for \textit{User 49} after instruction tuning, but without reinforcement learning (RL) tuning, contains several irrelevant keywords related to the interacted items, such as "Foodie," "Food and drink," and "Horoscope." Notably, these terms also appear in the profiles of \textit{User 11451} and \textit{User 20522}, suggesting that the generated profile is overly influenced by too many collaborative users. In contrast, the profile generated for \textit{User 49} after RL tuning effectively preserves the preferences indicated in the interaction history while incorporating relevant implicit keywords from collaborative users. For example, the term "pop culture" is derived from \textit{User 341}'s profile. This approach provides precise and valuable additional information for modeling \textit{User 49}'s preferences. We attribute this improvement to our proposed RL-based personalized feature enhancement techniques, which effectively address the noise and over-smoothing issues that can arise during the instruction-tuning process.

\section{Related Work}
\label{app:relate}
\subsection{ID-based Recommender Systems}
In recommender systems, numerous collaborative filtering models have been proposed to map users and items into latent representations based on user/item IDs~\citep{koren2021advances, su2009survey}. These methods have evolved significantly, starting from early matrix factorization techniques, such as BiasMF~\citep{koren2009matrix}, to the introduction of Neural Collaborative Filtering (NCF) with the advent of neural networks~\citep{he2017neural}. Recently, advancements in Graph Neural Networks (GNNs) have opened promising avenues for constructing bipartite graphs based on user-item interaction history, allowing for the capture of high-order collaborative relationships. GNN-based methods, including NGCF~\citep{wang2019neural}, GCCF~\citep{chen2020revisiting}, and LightGCN~\citep{he2020lightgcn}, have demonstrated state-of-the-art performance, enhancing the effectiveness of recommendation.

Additionally, researchers have incorporated self-supervised learning (SSL) techniques as supplementary learning objectives to improve the robustness of recommenders and address challenges related to data sparsity and noise~\citep{yu2023self}. Contrastive learning (CL), a widely adopted SSL technique, has been effectively applied in CF research through approaches such as SGL~\citep{wu2021self}, SimGCL~\citep{yu2022graph}, NCL~\citep{lin2022improving}, and AdaGCL~\citep{jiang2023adaptive}. Despite these advancements, ID-based recommenders still face significant limitations, particularly in completely cold-start scenarios and in terms of model transferability~\citep{yuan2023go}.

\subsection{Large Language Models (LLMs) for Recommendation}
The application of large language models (LLMs) in recommender systems has garnered significant attention~\citep{fan2023recommender, lin2023can, liu2023pre}. Current approaches can be categorized into two main types. The first category includes methods such as P5~\citep{geng2022recommendation} and Chat-REC~\citep{gao2023chat}, which emphasize designing prompts aligned with recommendation tasks, utilizing the LLM directly as the inference model. The second category enhances existing recommenders by integrating LLMs while still relying on traditional collaborative filtering methods~\citep{wang2024large}. For instance, LLMRec~\citep{wei2024llmrec} strengthens the user-item interaction graph through LLM-based graph augmentation, while RLMRec~\citep{ren2023representation} combines LLM-enhanced text embeddings with GNN-based user/item representations. However, these approaches often lack fine-tuning tailored to specific recommendation tasks, primarily focusing on full-shot scenarios.

In contrast, our work introduces a novel instruction-tuning technique for an open-source LLM, allowing it to adapt to specific recommendation tasks and effectively capture collaborative information for profile generation. While methods like InstructRec~\citep{zhang2023recommendation} and TALLRec~\citep{bao2023tallrec} align LLM capabilities with recommendation tasks, they struggle with scalability due to instruction-question-answering prompts and exhibit poor generalization on sparse data. Our approach enhances the generalization ability of existing recommender systems in the face of data scarcity and noise, while maintaining efficiency in handling large-scale data in practical scenarios.

\section{Conclusion}\vspace{-0.1in}
\label{sec:conclusion}

In this work, by seamlessly integrating large language models with collaborative filtering, we have introduced a novel instruction-tuning paradigm that equips language models with the critical ability to capture complex user-item interactions and behavioral preferences. This model-agnostic approach can be easily plugged into existing recommender systems, dramatically enhancing their generalization capacity, particularly in data-scarce scenarios where traditional ID-based methods often struggle. The two key innovations of \model\ - the fusion of external features with collaborative patterns through specialized instruction tuning, and the reinforcement learning-based personalized feature enhancement framework - address the fundamental challenges of cold-start profiling and data noise inherent in recommendation tasks. Comprehensive evaluations demonstrate the significant advantages and practical compatibility of our approach with state-of-the-art recommender systems.

\clearpage

\bibliography{main}
\bibliographystyle{iclr2025_conference}

\clearpage
\label{sec:appendix}
\section{Appendix}
\label{app:ins}

\subsection{Dataset Descriptions}
\label{app:dataset}
The statistical characteristics of the three datasets used in our experiments are summarized in Table~\ref{tab:dataset}. Below, we provide detailed descriptions of each dataset:

\label{app:dataset}
\begin{table}[h]
    \caption{Statistics of the experimental datasets.}
    \label{tab:dataset}
    \centering
    \resizebox{0.5\linewidth}{!}{
        \begin{tabular}{l|c|c|c}
            \hline
             \textit{Statistics} & \textbf{MIND} & \textbf{Netflix} & \textbf{Industrial} \\
             \hline
             \textbf{\# User} & 57128 & 16835 & 117433 \\
             \textbf{\# Overlap. Item} & 1020 & 6232 & 72417\\
             \textbf{\# Snapshot} & daily & yearly & daily \\
             \hline
             \multicolumn{4}{c}{Training Set} \\
             \hline
             \textbf{\# Item} & 2386 & 6532 & 152069 \\
             \textbf{\# Interactions} & 89734 & 1655395 & 858087 \\
             \textbf{\# Sparsity} & 99.934$\%$ & 98.495$\%$ & 99.995$\%$ \\
             \hline
             \multicolumn{4}{c}{Test Set} \\
             \hline
             \textbf{\# Item} & 2461 & 8413 & 158155 \\
             \textbf{\# Interactions} & 87974 & 1307051 & 876415 \\
             \textbf{\# Sparsity} & 99.937$\%$ & 99.077$\%$ & 99.995$\%$ \\
             \hline
        \end{tabular}
    }
\end{table}

\begin{itemize}[leftmargin=*]

\item \textbf{MIND Data}: A large-scale news recommendation benchmark dataset. We extracted data spanning two consecutive days, with one day allocated for training and the subsequent day for testing. Each news article contains rich textual features including category labels, titles, and abstract content.

\item \textbf{Netflix Data}: This dataset is derived from the Netflix Prize Data competition on Kaggle, comprising implicit user-movie interactions. For our experiments, we extracted viewing histories spanning two consecutive years (2005-2006), with the first year's data allocated for training and the subsequent year for testing. The textual features for each movie item are represented by its corresponding title.

\item \textbf{Industrial Data}: A large-scale proprietary dataset collected from a major online content platform (anonymized for privacy). This comprehensive news article collection serves millions of daily active users and generates substantial user-article interactions. For our evaluation, we extracted interaction data from two consecutive days (Day N and Day N+1), utilizing them as training and testing sets respectively. Each article is represented by its title as the primary textual feature.

\end{itemize}

\subsection{Base Model Descriptions}
\label{app:baseline}
We provide detailed descriptions of the baseline methods employed in our comparative evaluation:

\begin{itemize}[leftmargin=*]

\item \textbf{BiasMF}~\citep{koren2009matrix}: A matrix factorization-based approach that enhances recommendation accuracy by incorporating user and item bias vectors to capture individual preference patterns.

\item \textbf{NCF}~\citep{he2017neural}: A neural collaborative filtering framework that replaces traditional matrix factorization's dot-product operation with multi-layer neural networks, enabling the capture of complex user-item interactions. In our implementation, we specifically utilize the NeuMF variant.

\item \textbf{LightGCN}~\citep{he2020lightgcn}: A graph-based recommendation model that leverages user-item interaction graphs through a simplified layer-wise propagation scheme, employing only linear transformations and element-wise operations to aggregate neighborhood information.

\item \textbf{SGL}~\citep{wu2021self}: An enhanced version of LightGCN that incorporates self-supervised contrastive learning. It employs various data augmentation techniques, including random walks and node/edge dropout, to generate diverse graph views for contrastive learning.

\item \textbf{SimGCL}~\citep{yu2022graph}: A streamlined contrastive learning framework that generates contrastive views by directly injecting uniform noise into the embedding space, eliminating the need for explicit graph augmentation operations.

\end{itemize}

\subsection{Experimental Settings for Performance Comparison}
\label{app:overall}

We conducted our performance evaluation (detailed in Sec.~\ref{sec:rq1}) under two distinct testing scenarios:

\begin{itemize}[leftmargin=*]

\item \textbf{Full-Shot Setting}: Utilizes the complete test set, including items that appeared in the training phase, representing conventional recommendation scenarios.

\item \textbf{Zero-Shot Setting}: Exclusively evaluates items absent from the training set, specifically designed to assess cold-start recommendation capability where items have no historical interaction data.

\end{itemize}

Our experimental framework compares two implementation variants:

\begin{itemize}[leftmargin=*]

\item \textbf{Base Variant}: Implements our cold-start recommendation paradigm using only user ID embeddings and item text embeddings, without LLM-generated profiles.

\item \textbf{Augmented Variant}: Fully integrates the proposed \model\ framework with traditional recommenders, incorporating LLM-generated item profiles.

\end{itemize}

This comparative setup enables systematic evaluation of how LLM-generated profiles enhance recommender performance, particularly in addressing cold-start challenges.

\subsection{Ablation Study Configuration}
\label{app:ablation}

We evaluate four distinct variants of our model to assess the contribution of each component:

\begin{itemize}[leftmargin=*]

\item \textbf{GPT\_KD Variant}: Implements basic large language model (LLM) fine-tuning using ChatGPT3.5-generated user profiles, as detailed in Sec.~\ref{sec:gpt_kd tuning}.

\item \textbf{Naive Variant}: Builds upon GPT\_KD by incorporating instruction tuning, but restricts the dialogue to single-turn interactions, diverging from our proposed two-turn approach (Sec.~\ref{sec:two_turn tuning}).

\item \textbf{Mask Variant}: Extends GPT\_KD with both two-turn instruction tuning and masking-based strategies, representing an intermediate step in our framework.

\item \textbf{Ours (\model)}: Represents our complete framework, enhancing the Mask variant with RL-based personalized feature optimization.

\end{itemize}

\subsection{Implementation Details}
\label{app:imp}
\subsubsection{Parameter-Efficient Fine-Tuning Strategy}
\label{app:peft}
To optimize LLM adaptation while preserving their core reasoning capabilities, we employ Parameter-Efficient Fine-Tuning (PEFT) methodology. Specifically, we utilize Low-Rank Adaptation (LoRA)~\citep{hu2021lora} for fine-tuning Llama2-7b-chat~\citep{touvron2023llama}, enabling efficient task-specific adaptation while maintaining the model's fundamental knowledge base.

\subsubsection{Base Recommender Integration and Hyperparameter Configuration}
We systematically integrated \model\ into various base recommenders, conducting comprehensive hyperparameter optimization to ensure fair comparison. All models are implemented using PyTorch framework, utilizing Adam optimizer with default parameters and Xavier initialization. The training process employs a batch size of 4096, with embedding vectors dimensionality set to 32 and a learning rate of $1e^{-3}$. The $\mathcal{L}_2$ regularization coefficient is optimized through a search space of \{$1e^{-3}$, $1e^{-4}$, $1e^{-5}$, $1e^{-6}$, $1e^{-7}$\}. For GNN-based models (LightGCN, SGL, and SimGCL), we set the number of GCN layers to 2. Additionally, for self-supervised learning (SSL)-based models (SGL and SimGCL), the temperature coefficient is tuned within the range of $\{0.1, 0.5, 1.0\}$.

\subsection{Instruction Designs}
\label{app:ins}
In this section, we provide a comprehensive overview of the instructions utilized for fine-tuning at each stage of our process. We will also discuss the methodologies employed to construct both positive and negative training samples for the reinforcement learning reward model.

\begin{itemize}[leftmargin=*]

\item \textbf{Instruction Generation}. As illustrated in Figure~\ref{fig:ins_kd}, our instruction generation is conducted based on the knowledge distillation process through leveraging ChatGPT's capabilities by feeding it two key inputs: user-specific textual information and interaction-related item descriptions. The LLMs then synthesize this information to generate comprehensive user profiles that capture both personal identity characteristics and specific interest patterns.

    \begin{figure*}[h]
    \centering
    \includegraphics[width=1 \linewidth]{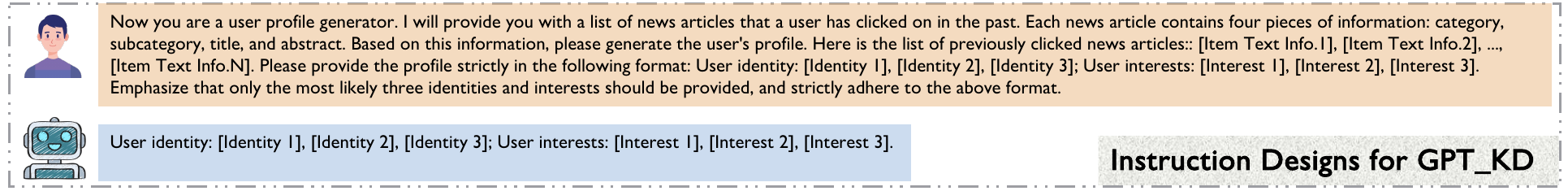}
    \caption{Instruction generation via ChatGPT knowledge distillation.}
    \label{fig:ins_kd}
    \end{figure*}

\item \textbf{Two-Turn Collaborative Instruction Design.} Our instruction tuning framework implements a carefully structured two-turn dialogue approach (Figure~\ref{fig:ins_two_turn}). \textbf{Initial Setup:} The system receives foundational instructions to establish LLMs' understanding of collaborative filtering principles.

    \textbf{First Turn:} In the first turn, we provide input instructions containing interaction histories from similar users and relevant item details. These similar users are identified through a graph collaborative filtering model, followed by embedding-based similarity calculations. The LLMs then process this information to generate comprehensive user profiles for each referenced user.

    \textbf{Second Turn:} In the second turn, the LLMs perform prediction tasks by analyzing the previously generated user profiles in conjunction with specific item characteristics. Based on collaborative filtering principles, the model evaluates potential user-item interactions and outputs definitive binary predictions (Yes/No) regarding interaction likelihood.
    
    \begin{figure*}[h]
    \centering
    \includegraphics[width=1 \linewidth]{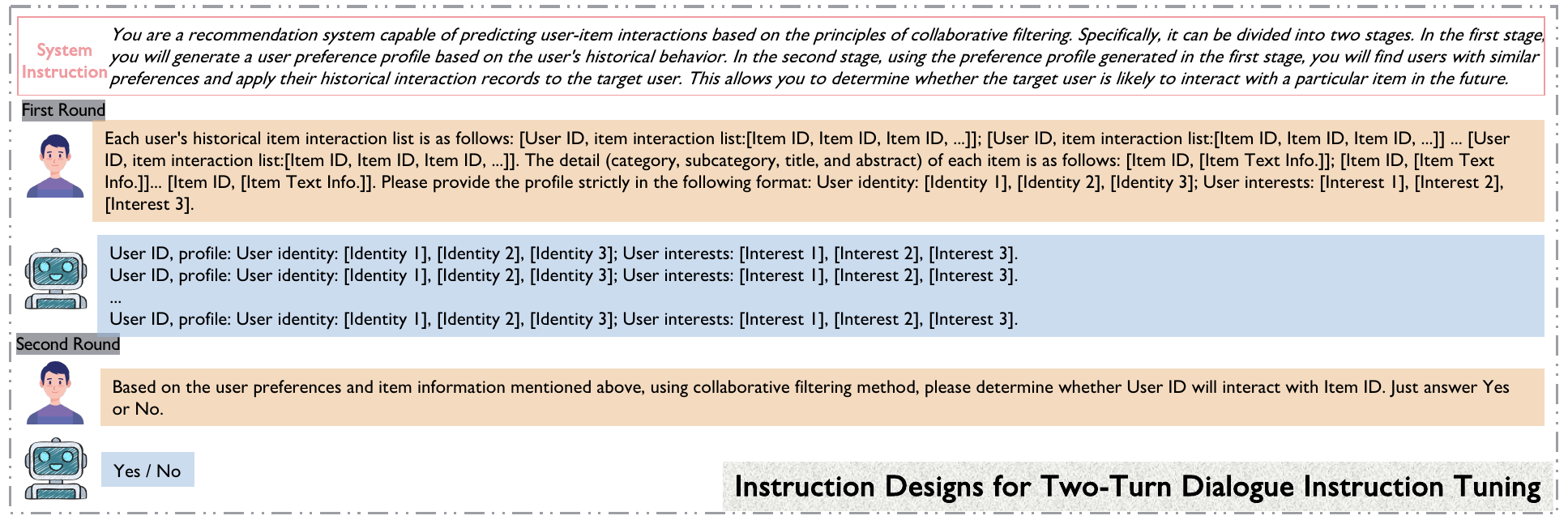}
    \caption{Instruction designs for two-turn instruction tuning.}
    \label{fig:ins_two_turn}
    \end{figure*}

 \item \textbf{User Profile Generation Instruction Design.} Following the instruction tuning, the LLMs develop proficiency in profile generation with collaborative awareness. As illustrated in Figure~\ref{fig:ins_user_profile}, we implement a systematic instruction framework for user profile generation that consists of three key components. The process begins with system-level instructions that reinforce the LLMs' understanding of collaborative filtering principles. The input layer then incorporates two essential elements: (1) interaction histories from multiple similar users, including the target user requiring profile generation, and (2) comprehensive textual descriptions of all relevant items. Based on this information, the LLMs synthesize and output a detailed profile for the target user.

    \begin{figure*}[h]
    \centering
    \includegraphics[width=1 \linewidth]{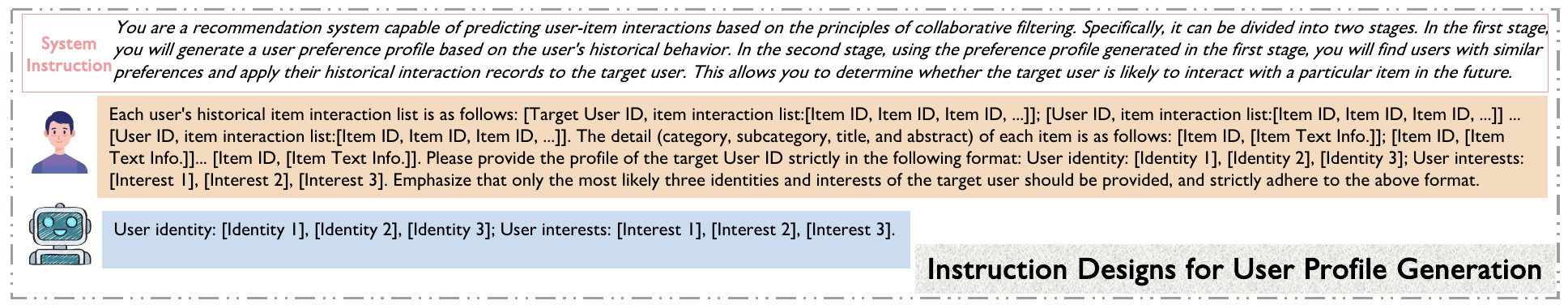}
    \caption{Instruction designs for user profile generation.}
    \label{fig:ins_user_profile}
    \end{figure*}

 \item \textbf{Item Profile Generation Instruction Design.} Following user profile generation, we develop item profiles to ensure semantic alignment between user and item features. The item profile represents a target user's profile specific to a particular item, and we implement a dual-phase approach for comprehensive coverage: In the first phase, we generate profiles for items with existing user interactions by leveraging the profiles of their interacting users. This establishes direct user-item semantic connections. For cold-start items, the second phase employs similarity-based matching using raw item embeddings, followed by LLM-based profile inference. As illustrated in Figure~\ref{fig:ins_item_profile}, the instruction framework processes inputs comprising: (1) a target item and its similar items, (2) detailed textual information for all items, and (3) existing profiles of similar items. The LLMs then synthesize this information to generate the target item's profile, thereby enhancing the overall semantic coherence of the recommendation system.

    \begin{figure*}[h]
    \centering
    \includegraphics[width=1 \linewidth]{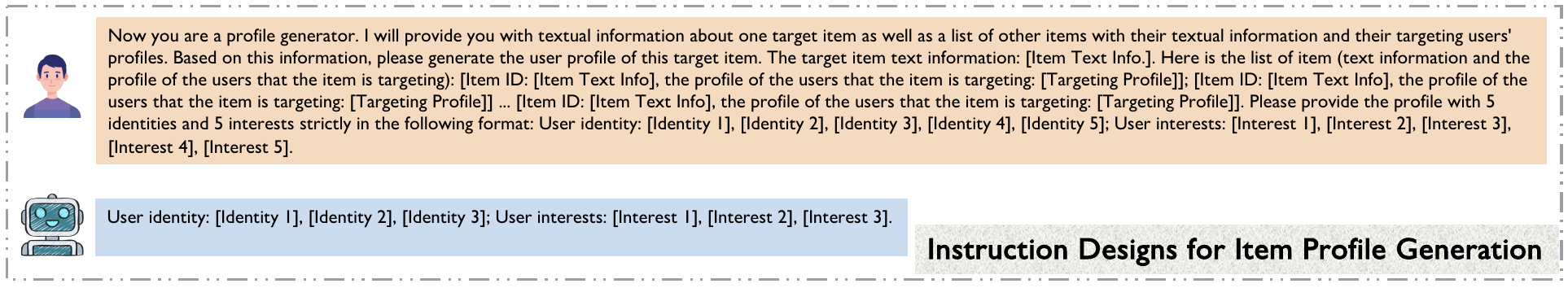}
    \caption{Instruction designs for item profile generation.}
    \label{fig:ins_item_profile}
    \end{figure*}

\item \textbf{Reward Model Training: Response Construction.} The success of our personalized feature enhancement through reinforcement learning (Sec~\ref{subsec:RL_llm}) depends critically on the quality of the reward model training data. This training data addresses two key challenges: instruction-tuning noise and collaborative feature over-smoothing. As illustrated in Figure~\ref{fig:ins_rm}, we construct positive samples through a combination of state-of-the-art LLMs (\textit{e.g.}, ChatGPT) and manual validation. For negative samples, we establish a complementary sampling strategy:
\begin{itemize}[leftmargin=*]
    \item \textbf{Similar User Profiles:} These samples train the reward model to identify subtle profile distinctions, thereby mitigating over-smoothing effects.
    \item \textbf{Low-Quality Responses:} This category includes profiles with missing information or repetitive content, serving as explicit negative examples for model training.
\end{itemize}

    \begin{figure*}[h]
    \centering
    \includegraphics[width=1 \linewidth]{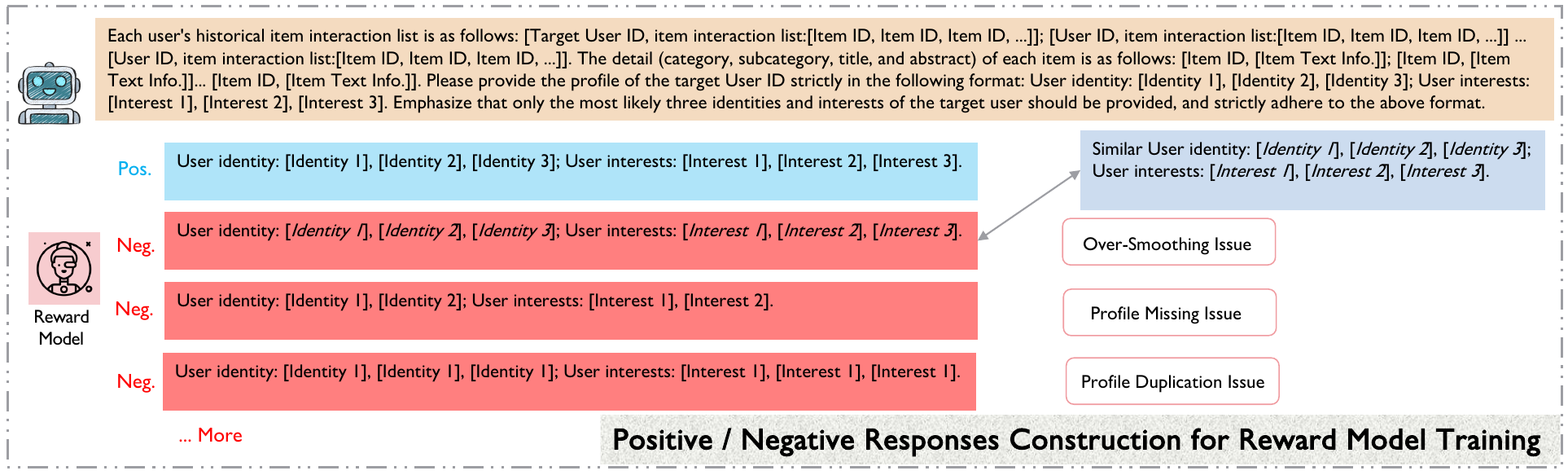}
    \caption{Positive/Negative responses construction for reward model training.}
    \label{fig:ins_rm}
    \end{figure*}
    
\end{itemize}

\subsection{Future Work Exploration}
Future research directions for RecLM could explore two promising avenues: First, extending the framework to multi-modal contexts, where user-item interactions involve diverse data types such as images, videos, and audio. This would require developing modal-specific instruction tuning strategies and designing cross-modal alignment mechanisms to maintain semantic consistency across different modalities. Second, investigating the potential of smaller, more efficient language models for profile generation. While our current implementation leverages Llama2-7b, exploring distilled or compressed models could significantly reduce computational overhead while maintaining profile quality. This could involve techniques like knowledge distillation from larger to smaller models specifically trained for recommendation tasks, or developing specialized lightweight architectures that focus exclusively on user/item profiling capabilities.

\end{document}